\documentclass[a4paper,10pt,twocolumn]{article}

\usepackage[utf8]{inputenc}
\usepackage[T1]{fontenc}
\usepackage{amsmath, amssymb, amsxtra}
\usepackage{nccmath, icomma, xspace, bbold}
\usepackage{cite}
\usepackage{siunitx}
\DeclareSIUnit\dbkm{dB/km}
\DeclareSIUnit\krec{\ensuremath{\textit{k}_\text{rec}}}
\DeclareSIUnit\Erec{\ensuremath{\textit{E}_\text{rec}}}
\usepackage[normalem]{ulem}
\usepackage{color}

\usepackage[%
	left = 1.5cm,%
	right = 1.5cm,%
	top = 1cm,%
	bottom = 1.5cm,%
	footskip = 0.5cm,%
	nohead]
	{geometry}
\usepackage{nccparskip}
\SetParskip{4pt}
\usepackage{setspace}
\usepackage{enumitem}

\usepackage[british]{babel}
\usepackage{csquotes}

\usepackage{ctable}
\usepackage{caption}%
\DeclareCaptionLabelSeparator{vertline}{~|~}%
\usepackage{authblk}

\usepackage{siunitx}
\usepackage{subcaption}

\usepackage{xifthen}
\newcommand{\ket}[2][]{ %
	\ifthenelse{\isempty{#1}}%
	{\ensuremath{\xspace\left\vert #2 \right\rangle\xspace}}%
	{\ensuremath{\xspace\left\vert #2 \right\rangle_{\! #1}\xspace}}
	}
\newcommand{\bra}[2][]{ %
	\ifthenelse{\isempty{#1}}%
	{\ensuremath{\xspace\left\langle #2 \right\vert\xspace}}
	{\ensuremath{\xspace\prescript{}{#1}{\!\left\langle #2 \right\vert\xspace}}}
	}
\newcommand{\braket}[3][]{ %
	\ifthenelse{\isempty{#1}}%
	{\ensuremath{\xspace\left\langle #2 \left\vert\right. #3 \right\rangle\xspace}}
	{\ensuremath{\xspace\left\langle #2 \left\vert\right. #3 \right\rangle_{\! #1}\xspace}}
	}
\newcommand{\avg}[2][]{ %
	\ifthenelse{\isempty{#1}}%
	{\ensuremath{\xspace\langle #2 \rangle\xspace}}%
	{\ensuremath{\xspace\langle #2 \rangle_{\! #1}\xspace}}
	}

\newcommand{\krec}{\ensuremath{k_\text{rec}}\xspace}
\newcommand{\Erec}{\ensuremath{E_\text{rec}}\xspace}

\begin{document}
\title{\bf
Long-distance single photon transmission from a trapped ion via quantum frequency conversion}
\author[1]{Thomas Walker}
\author[2]{Koichiro Miyanishi}
\author[2]{Rikizo Ikuta}
\author[1]{Hiroki Takahashi} 
\author[1]{Samir Vartabi Kashanian}
\author[3]{Yoshiaki Tsujimoto}
\author[3]{Kazuhiro Hayasaka}
\author[2]{Takashi Yamamoto}
\author[2]{Nobuyuki Imoto}
\author[1]{Matthias Keller}

\affil[1]{Department of Physics and Astronomy, University of Sussex, Brighton, BN1 9RH, United Kingdom}
\affil[2]{Graduate School of Engineering Science, Osaka University, Toyonaka, Osaka 560-8531, Japan}
\affil[3]{Advanced ICT Research Institute, National Institute of Information and Communications Technology (NICT),Koganei, Tokyo 184-8795, Japan}

\date{}
\maketitle

\textbf{
Trapped atomic ions are ideal single photon emitters with long lived internal states which can be entangled with emitted photons \cite{monroe2002}.  Coupling the ion to an optical cavity enables efficient emission of single photons into a single spatial mode and  grants control over their temporal shape \cite{keller2004}. 
These features are key for quantum information processing\cite{moehring07,nickerson2014} and quantum communication \cite{gisin2007,simon2017}.
However, the photons emitted by these systems are unsuitable for long-distance transmission due to their wavelengths. Here we report the transmission of single photons from a single $^{40}\text{Ca}^{+}$ ion coupled to an optical cavity over a \SI{10}{\km} optical fibre via frequency conversion from \SI{866}{\nm} to the telecom C-band at \SI{1530}{\nm}. We observe non-classical photon statistics of the direct cavity emission, the converted photons and the \SI{10}{\km} transmitted photons, as well as the preservation of the photons' temporal shape throughout. This telecommunication ready system can be a key component for long-distance quantum communication as well as future cloud quantum computation \cite{barz2012,fitzsimons2017}.
}

Accessibility of matter quantum systems to standard telecommunication wavelengths is a prerequisite for long-distance optical fibre quantum networks. The size of the network can be expanded by quantum repeater algorithms where matter systems play an important role in storing and providing non-classical states of light \cite{briegel1998}. Trapped ions coupled to optical cavities as individual qubit systems have been used to demonstrate single photon emission with controlled temporal shapes \cite{keller2004} and polarisation \cite{barros2009}, and the entanglement of ions and photons \cite{stute2012}.
Until recently, however, the emission wavelength was limited by the internal structure of the ions. Often in the visible or near-infra-red (NIR) regions, use of this emission is impractical over long distances due to the high attenuation in optical fibres. Further, communicaiton between disparate quantum systems was impossible due to their mismatched emission and absorption wavelengths. Quantum frequency conversion (QFC), which shifts the frequency of a photon without disturbing its quantum properties, is a solution to these issues. Since the first proposal of QFC \cite{kumar1990}, conversion between a wide range of frequencies have been demonstrated. Initially, this was used for detecting telecom photons with silicon photon detectors, which are usually confined to the visible and NIR regions \cite{langrock2005}. Quantum state preservation has been demonstrated in conversion between the NIR and telecom bands \cite{tanzilli2005,ikuta2011,ikuta2013high}. Furthermore, it has been applied to semiconductor quantum dots \cite{rakher2010,de2012}, and atomic ensembles \cite{radnaev2010,dudin2010,farrera2016,ikuta2016}. Recently, a QFC system was employed to generate 
entanglement between a trapped ion and a frequency converted photon\cite{bock2017}.
In contrast to systems based on spontaneous emission, the emission of trapped ions coupled to an optical cavity can be coherently controlled and provides highly efficient emission into a single spatial mode.  The ability to control the temporal shape of emitted photons is vital for a quantum interface to transfer entanglement between different systems. In this letter, we demonstrate QFC of photons from a single ion coupled to an optical cavity from \SI{866}{\nm} to \SI{1530}{\nm}, and the transmission of the converted photons through a \SI{10}{\km} optical fibre.

\begin{figure*}[t]
\begin{center}

\includegraphics[width=\linewidth]{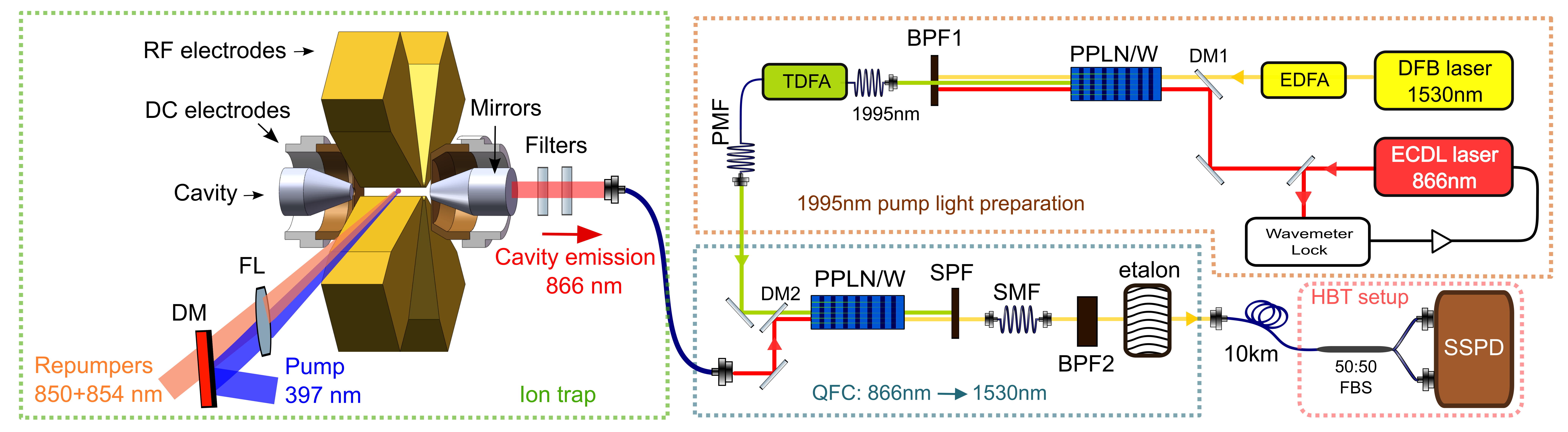}
\caption{\textbf{The experimental set-up.} The ion trap setup is shown on the left. Rf and DC electrodes provide radial and axial confinement of the single ion. Lasers for cooling, repumping, and single photon generation are overlapped at a dichroic mirror (DM) and focused to the trap centre by a lens (FL). The cavity length is stabilised via a reference laser locked to the D$_1$ transition of Cs at \SI{894}{\nm}. After a pair of longpass filters attenuate the \SI{894}{\nm} locking light to within the background rate of the detectors, the \SI{866}{\nm} cavity emission is coupled into a single mode optical fibre, after which it may be coupled to a Hanbury-Brown-Twiss (HBT) set-up for \SI{866}{\nm} photons, or to the QFC set-up as shown. The optical circuit for QFC is composed of two blocks: \SI{1995}{\nm} pump light preparation and QFC from \SI{866}{\nm} to \SI{1530}{\nm}. This is in turn connected to a HBT set-up either directly or via a \SI{10}{km} optical fibre.  \label{fig:setup}}
\end{center}
\end{figure*}

\begin{figure}[h]
\begin{center}
 \includegraphics[width=\linewidth]{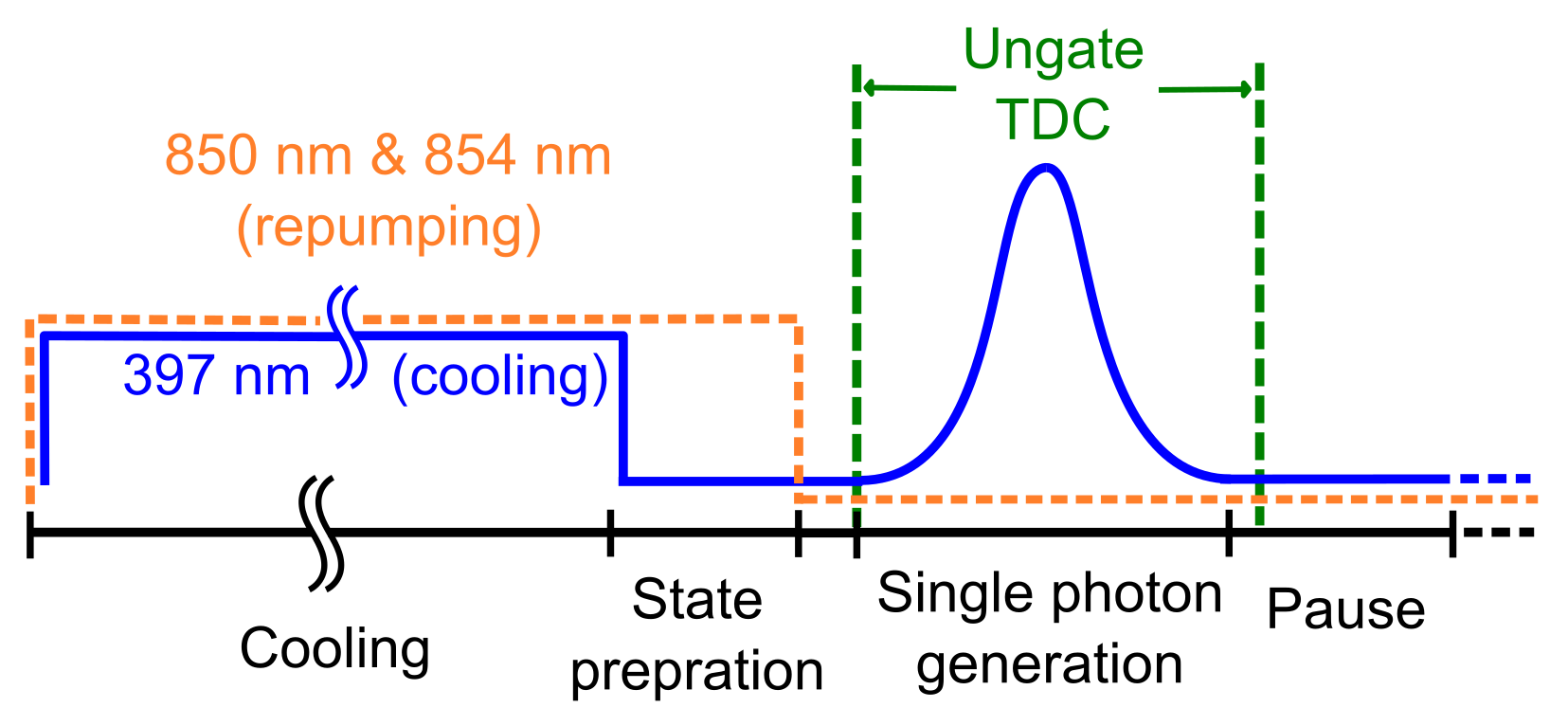}
 \caption{\textbf{The single photon generation sequence.} This figure shows the sequence and shapes of laser pulses used to generate single photons (not to scale). The ion is initially cooled for \SI{5.5}{\us}, and then pumped into the S$_{1/2}$ state for \SI{1}{\us}. After a brief delay, a Gaussian-shaped pulse drives a cavity assisted Raman transition to produce a single photon, during which time the time-to-digital converter (TDC) records photon arrive times. The sequence is repeated at 99.6 kHz.
   \label{fig:seq}}
   \end{center}
\end{figure}

\textbf{The ion-cavity set-up.}
The ion trap is similar to that described in \cite{begley2016} and is shown in Figure \ref{fig:setup}. $^{40}\text{Ca}^{+}$ ions are trapped in a linear Paul trap, in which symmetric radio frequency (rf) voltages on four blade-shaped electrodes form the radial trapping potential with a secular frequency of \SI{1.23}{\MHz}, and DC voltages on two endcap electrodes provide axial confinement with a secular frequency of \SI{760}{kHz}. The distance between the tips of the rf electrodes and trap centre is \SI{475}{\um}, and the DC electrodes are separated by \SI{5}{\mm}. The ion is coupled to an optical cavity formed by two mirrors, which are placed inside the endcap electrodes such that the cavity axis is collinear with the trap axis. One mirror, with a transmissivity of \SI{100}{ppm} at \SI{866}{\nm} acts as an output coupler, with the second mirror at \SI{5}{ppm} leading to a finesse of 60,000. The cavity length is \SI{5.3}{\mm}, leading to ion-cavity coupling strength of $g_0 = 2\pi \times \SI{0.9}{\MHz}$.  A weak magnetic field of \SI{0.5}{gauss} directed along the cavity axis defines the quantization axis for the ion without causing significant Zeeman splitting.

The single photons were generated in a \SI{10}{\micro\second} sequence as shown in Figure \ref{fig:seq}. First, the ion is Doppler cooled using a laser red-detuned by $\frac{\Gamma}{2}$ ($\Gamma = 2\pi \times \SI{22}{\MHz}$) from resonance with the $^2$S$_{1/2}-^2$P$_{1/2}$ transition at \SI{397}{\nm}. Lasers at \SI{850}{\nm} and \SI{854}{\nm} are used to repump the ion out of the metastable $^2$D$_{3/2}$-state back into the cooling cycle. After cooling for \SI{5.5}{\micro\second}, the cooling beam is switched off for \SI{1}{\micro\second} to prepare the ion in the S$_{1/2}$-state. Then, a pulse of \SI{397}{\nm} light with a Gaussian intensity profile drives a cavity assisted Raman transition between the $^2$S$_{1/2}$ and $^2$D$_{3/2}$ states, producing a \SI{866}{\nm} photon in the caivty (similar to \cite{keller2004}). The photon emitted from the cavity is will be circularly polarised with a roughly equal probability of left or right handedness. The \SI{1}{\s} lifetime of the $^2$D$_{3/2}$ state ensures that only a single photon may be produced. A brief delay before and after the single photon drive pulse ensures the complete switch-off of all the lasers used for cooling and state preparation to prevent repopulation of the $^2$S$_{1/2}$ and the generation of multiple photons. While increasing the single photon pulse length increases efficiency \cite{vasilev2010}, it also increases the amount of noise photons collected. To maximise signal-to-background ratio (SBR), a length of \SI{2.3}{\micro\second} was chosen and used throughout all measurements, as beyond this length the increase is noise outweighed the increase in efficiency. Using this scheme a single photon detection probability of $(1.63 \pm 0.04)\%$ is achieved with the superconducting single photon detectors as described later.

\textbf{The quantum frequency converter.}
To convert the single photons from the ion to the telecom C-band, a QFC set-up was built and connected between the trap and a Hanbury-Brown-Twiss (HBT) set-up for \SI{1530}{\nm} light. The QFC system uses difference frequency generation (DFG) between the photons at \SI{866}{\nm} and strong \SI{1995}{\nm} pump light to generate \SI{1530}{\nm} photons. The full quantum frequency conversion set-up is shown in Figure \ref{fig:setup}, which is composed of two optical circuits: one for \SI{1995}{\nm} pump light preparation and one for QFC of the \SI{866}{\nm} photons to \SI{1530}{\nm}. At the pump light preparation stage, the \SI{1530}{\nm} light from a distributed feedback (DFB) laser with a linewidth of \SI{10}{\MHz} 
is amplified by an erbium-doped fibre amplifier (EDFA) to a maximal power of \SI{450}{\mW}. The light at \SI{866}{\nm} from an external cavity diode laser (ECDL) is locked at a frequency resonant with the transition of ${}^{2}{\rm D}_{3/2}-{}^{2}{\rm P}_{1/2}$ of ${\rm Ca}^+$ using a wavemeter. The \SI{866}{\nm} and \SI{1530}{\nm} lasers are combined at a dichroic mirror~(DM1) and coupled into a periodically poled lithium niobate waveguide~(PPLN/W). \SI{1995}{\nm} light is prepared via DFG of \SI{866}{\nm} light with strong pump light at \SI{1530}{\nm}. 
Using a band pass filter~(BPF1) with a centre wavelength at \SI{2000}{\nm} and bandwidth of \SI{50}{\nm}, the \SI{1530}{\nm} pump light 
and the remaining seed light at \SI{866}{\nm} are filtered out from the output. The extracted \SI{1995}{\nm} light is coupled to a polarisation maintaining fibre~(PMF),
and is amplified by a thulium-doped fibre amplifier~(TDFA) to a maximum power of \SI{1}{\W}. 

At the QFC stage, the amplified vertically (V) polarised \SI{1995}{\nm} pump light is injected to another PPLN/W. 
The \SI{866}{\nm} photons from the ion-cavity system are combined with the pump by a dichroic mirror~(DM2) and coupled to the PPLN/W. 
After the PPLN/W, the strong pump is eliminated by a short pass filter~(SPF), and the converted \SI{1530}{\nm} photons 
are coupled into a single-mode fibre~(SMF). The SMF is connected to a frequency filter circuit, consisting of a fibre-coupled band pass filter~(BPF2) with a bandwidth of \SI{0.2}{\nm} and an etalon with a FWHM of $\sim$\SI{700}{\MHz} for suppressing noise photons picked up through the QFC system. After the filters, the light is coupled to a SMF. 

\begin{figure}[t]
 \begin{center}
 \includegraphics[width=\linewidth]{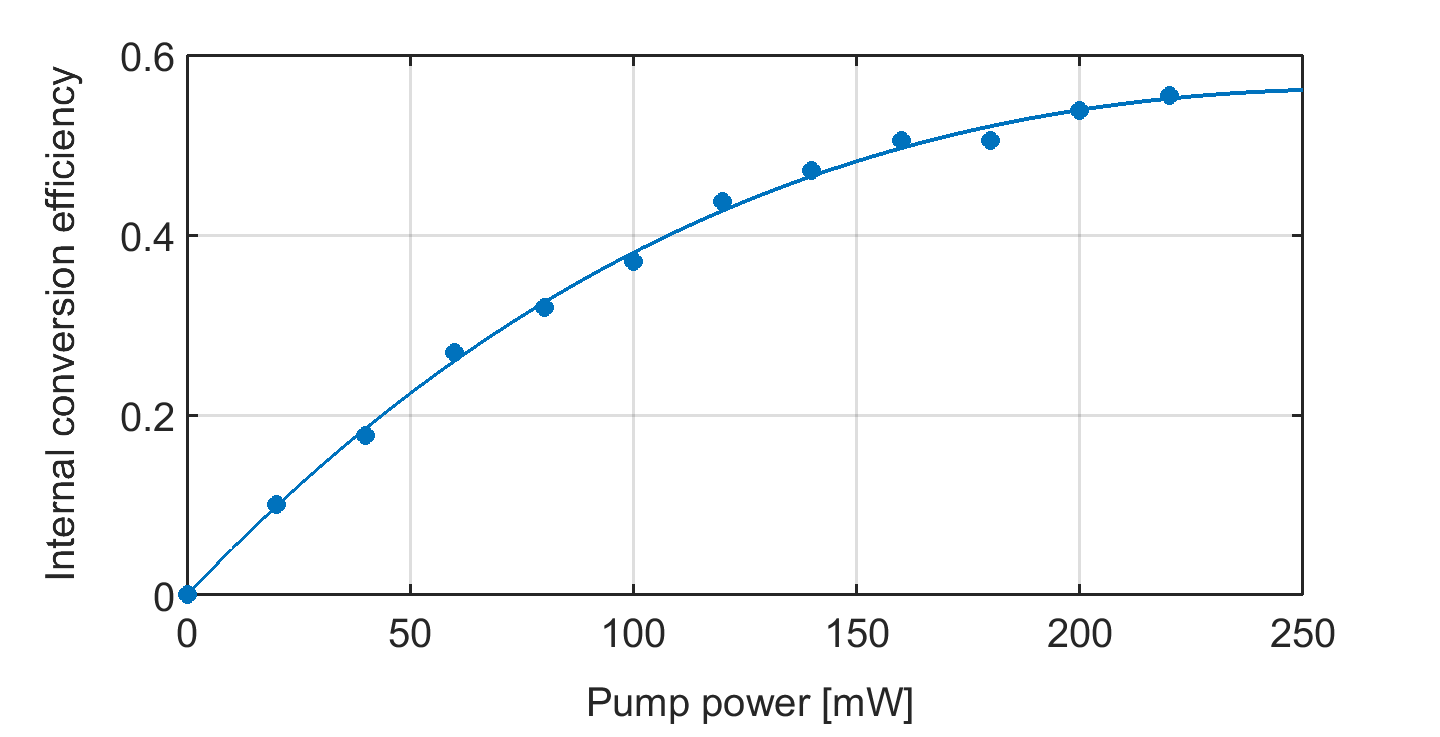}
 \caption{The internal conversion efficiency from \SI{866}{\nm} to \SI{1530}{\nm}.
   A solid curve is obtained by the best fit to the experimental data
   with a function $A\sin^2(\sqrt{B P})$, where $P$ is
   the power of the pump light at \SI{1995}{\nm},
   and the fitting parameters $A$ and $B$ are 0.56 and 9.3/W, respectively. 
   \label{fig:efficiency}}
 \end{center}
\end{figure}

The conversion efficiency of the QFC system was characterised using V-polarised laser light at \SI{866}{\nm} from the ECDL. The coupling efficiency of the \SI{866}{\nm} light to the PPLN/W is 80\%. 
The conversion efficiency from \SI{866}{\nm} light to \SI{1530}{\nm} depends on the pump power used, as shown in Figure \ref{fig:efficiency}.
The maximum conversion efficiency is about 50\% at \SI{0.2}{\W} effective pump power. 
The coupling efficiency of the converted light to the SMF is 60\%. The peak transmittance of the frequency filters (BFP2 and the etalon) is 
14\% including fibre coupling. 
A pump power of \SI{170}{\mW} was used for the experiment, for which the internal conversion efficiency is $\sim 50$\%. 
As a result, the total efficiency is about 3\%.

\begin{figure}[!h]
    \begin{center}

    \begin{subfigure}[t]{0.9\linewidth}
        \centering
        \includegraphics[width=\linewidth]{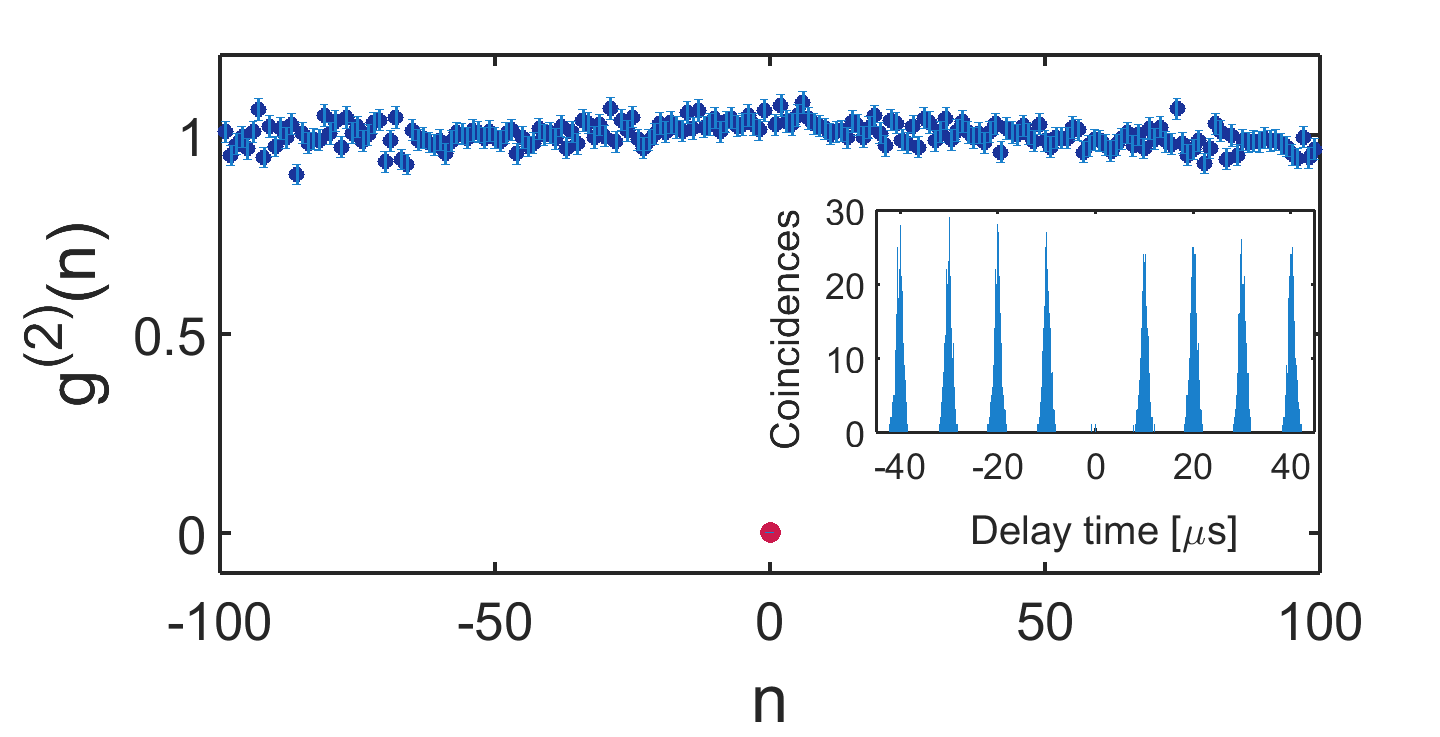}
        \caption{\label{fig:g2866}}
    \end{subfigure}
    \\
    \begin{subfigure}[t]{0.9\linewidth}
        \centering
        \includegraphics[width=\linewidth]{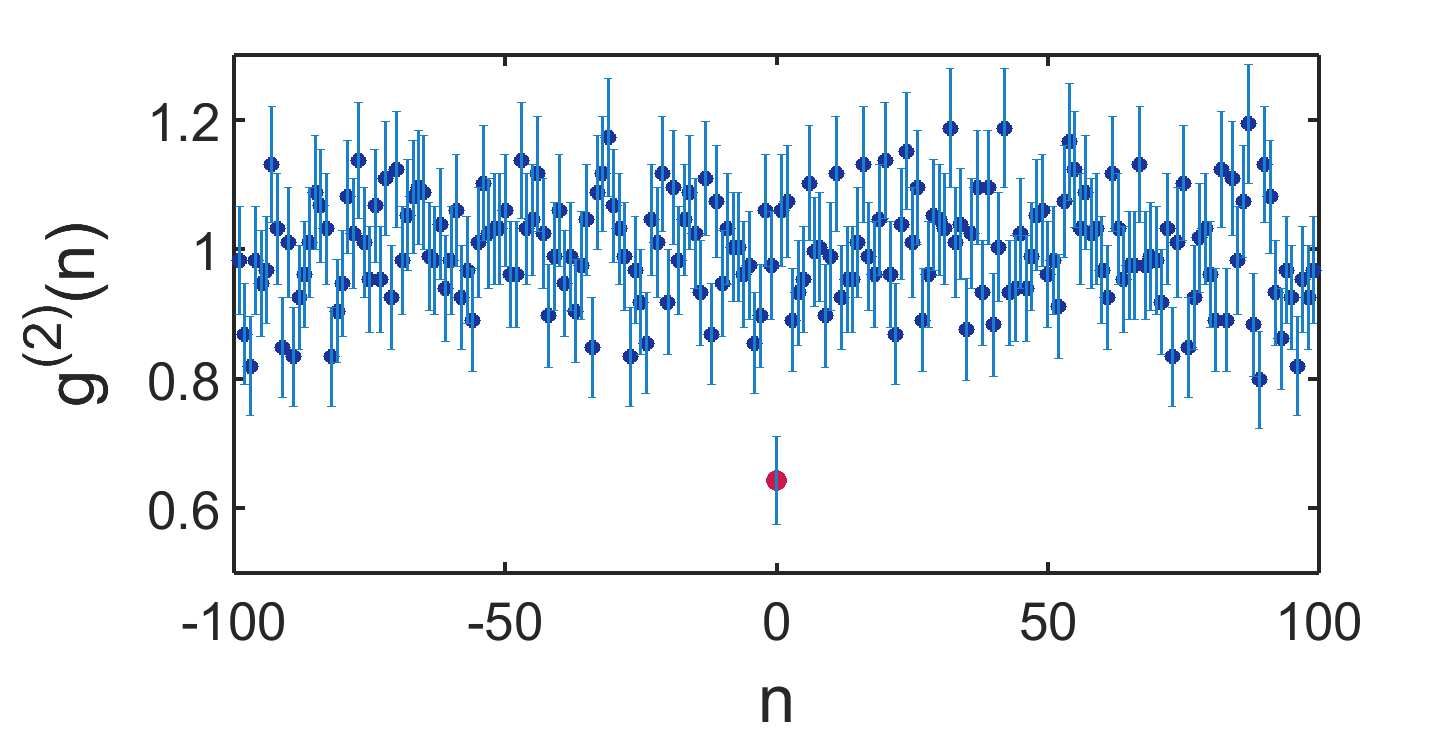}
        \caption{\label{fig:g21530}}
    \end{subfigure}
    \\
    \begin{subfigure}[t]{0.9\linewidth}
        \centering
        \includegraphics[width=\linewidth]{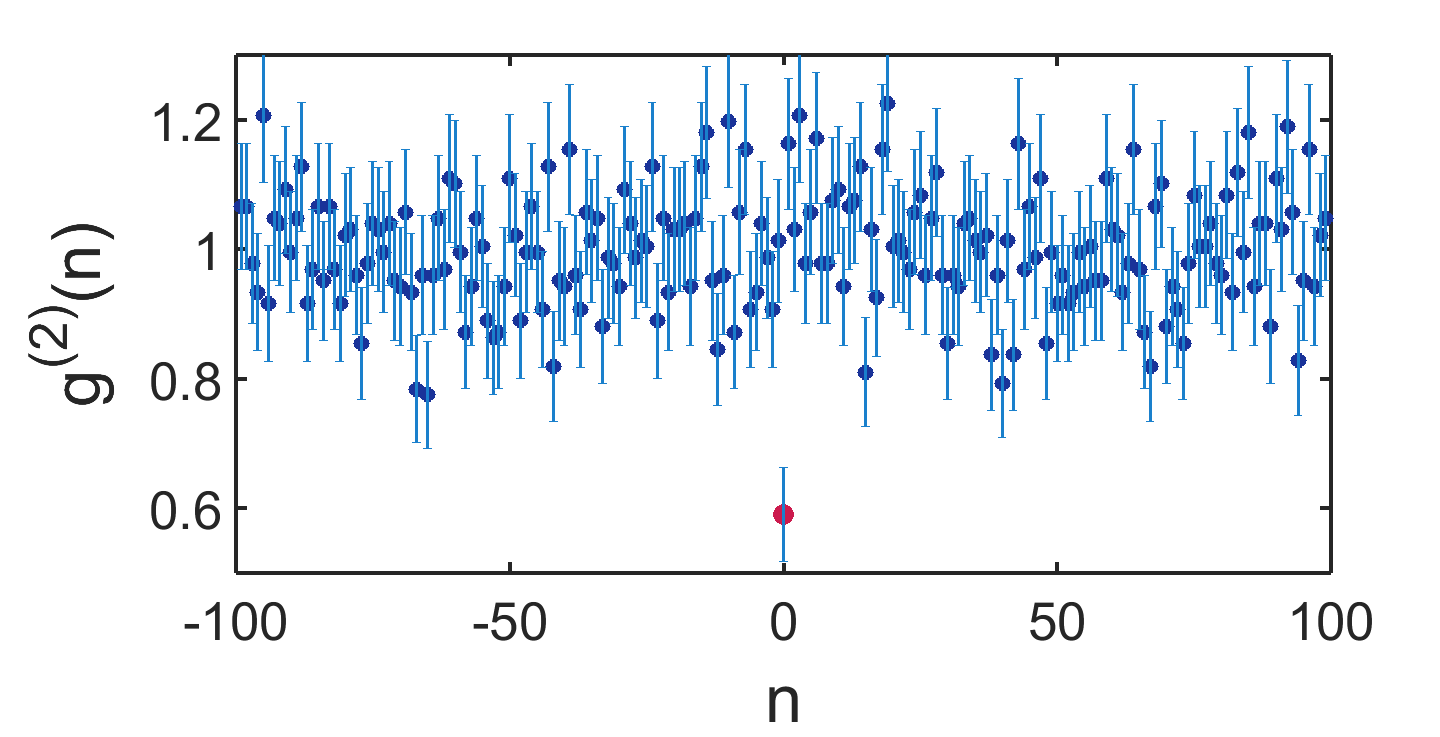}
        \caption{\label{fig:g210km}}
    \end{subfigure}
    \caption{The $g^{(2)}$ for delays up to \SI{1}{\ms} for \textbf{(a)} the \SI{866}{\nm} photons, \textbf{(b)} \SI{1530}{\nm} photons, and \textbf{(c)} \SI{1530}{\nm} photons with the \SI{10}{\km} fibre. The value at zero delay (shown in red) is clearly separated from the values at $n \neq 0$ in all cases, demonstrating sub-Poissonian statistics. The insert in \textbf{(a)} for the coincidence events at \SI{866}{\nm} up to \SI{50}{\us} in \SI{200}{\ns} bins. \label{fig:g2}}
    \end{center}
\end{figure}

\textbf{Characterisation of the single photons.} To measure the second order intensity correlation function of the single photons, $g^{(2)}$, a HBT set-up is employed. A fibre based 50/50 beam splitter (FBS) directs the cavity emission onto two superconducting single photon detectors (SSPDs) with a specified quantum efficiency of $> 80$\% at \SI{850}{\nm} (Single-Element Nanowire Detector from Photonspot Inc.). Similarly, after the QFC circuit, the \SI{1530}{\nm} photons are measured with a HBT set-up using SSPDs with a specified quantum efficiency of $> 80$\% at \SI{1550}{\nm}.

The electronic output pulses of the detectors are then recorded on two channels of a time-to-digital converter (TDC) (MCS6A from FastComtec), generating a list of photon arrival times paired with the corresponding channel numbers. The difference in arrival times of photons in each channel are sorted into \SI{200}{\ns} bins, resulting in a histogram as shown in the inset of Figure \ref{fig:g2866}. The series of regular sharp peaks with a suppressed peak at zero delay is characteristic of a pulsed single photon source, where the $n$th peak represents the number of coincidences between $n$th neighbour pulses. The peaks are separated by the period of the single photon sequence and have widths determined by the single photon shape.  We calculate $g^{(2)}$ from the total number of coincidence events in each peak, $N_{\text{coinc},n}$, the total number of counts in each channel, $N_1$ and $N_2$, and the number of times the sequence was repeated, $N_{\text{trig}}$, by

\begin{equation}
g^{(2)}(n) = \frac{N_{\text{trig}}N_{\text{coinc},n}}{N_1N_2},
\end{equation}
as in \cite{ikuta2011}. $N_{\text{trig}}$ is calculated as the total measurement time divided by the duration of a single cycle.

The data include not only correlations between single photon events, but also correlations between single photons and the background noise, as well as correlations within the noise. Therefore, we expect to measure a non-zero $g^{(2)}(0)$, with the measured value depending on the signal-to-background ratio. If the signal and background are statistically independent, and the intensity correlation of the background is 1, this relationship is given by

\begin{equation}
g^{(2)}(n) = 1 + \rho^2(g^{(2)}_{\text{signal}}(n)-1)
\end{equation}
where $\rho = \text{SBR}/(1+\text{SBR})$ \cite{Becher2001}. If a perfect single photon source is assumed (that is, $g^{(2)}_{\text{signal}}(0) = 0$) we have
\begin{equation}
g^{(2)}(0) = 1 - \rho^2.
\end{equation}

The measured $g^{(2)}(n)$ for the cavity emission at \SI{866}{\nm} is shown in Figure \ref{fig:g2866}. 294,020 single photon events were detected over 180 seconds: an average rate of \SI{1633}{counts\per\s}. In this time only 2 coincidence events were observed at $n=0$, giving a normalized $g^{(2)}(0) = 0.0017 \pm 0.0012$. The background count rate was measured by allowing the experimental sequence to run as normal with no ion present in the trap. In this way, a rate of \SI{1}{counts\per\s} was measured. Given this SBR, we expect $g^{(2)}(0) = 0.0012$; these coincidence events can therefore be fully accounted for by the background count rate.

The ion-cavity system was then connected to the QFC set-up. With QFC efficiency of 3\%, the measured single photon rate at \SI{1530}{\nm} was \SI{45.3}{counts\per\s}, with 1,325,433 total counts observed over 8.1 hours. Due to background light picked up through the QFC system, the gated background count rate was \SI{19}{counts\per\s} (again measured by running the sequence with no ion present), giving a SBR of 1.38. We observed $g^{(2)}(0) = 0.67 \pm 0.07$, as shown in Figure \ref{fig:g21530}, which is in agreement with the expected $g^{(2)}(0)$ value of $0.660$ for this SBR. The $g^{(2)}(0)$ is significantly below the classical limit and clearly demonstrates the conversion of single photons from the ion-cavity system from \SI{866}{\nm} to \SI{1530}{\nm}.

To demonstrate that the quantum nature of the single photons is preserved even over long distances, a \SI{10}{\km} SMF was inserted between the QFC set-up and the HBT set-up. The typical attenuation through an optical fibre at \SI{866}{\nm} is \SI{3}{\dbkm}; over \SI{10}{\km}, this equates to a reduction in signal by a factor of 1,000. At \SI{1530}{\nm}, however, the attenuation is only \SI{0.2}{\dbkm}: a 40\% reduction in signal over the same distance. After the \SI{10}{\km} fibre we observed an average rate of \SI{21.5}{counts\per\s}. However, the background rate was also attenuated to \SI{8.5}{counts\per\s}, resulting in an average SBR of 1.53, similar to the previous value. Using QFC we achieve a count rate an order of magnitude greater than would be possible with direct transmission of \SI{866}{\nm}. The data set shown in Figure \ref{fig:g210km} includes a total 2,081,939 counts collected over 26.9 hours. From this we extract $g^{(2)}(0) = 0.59 \pm 0.07$, in agreement with both the previous value and the calculated value of 0.634.

\begin{figure}[t]
 \begin{center}
 \includegraphics[width=\linewidth]{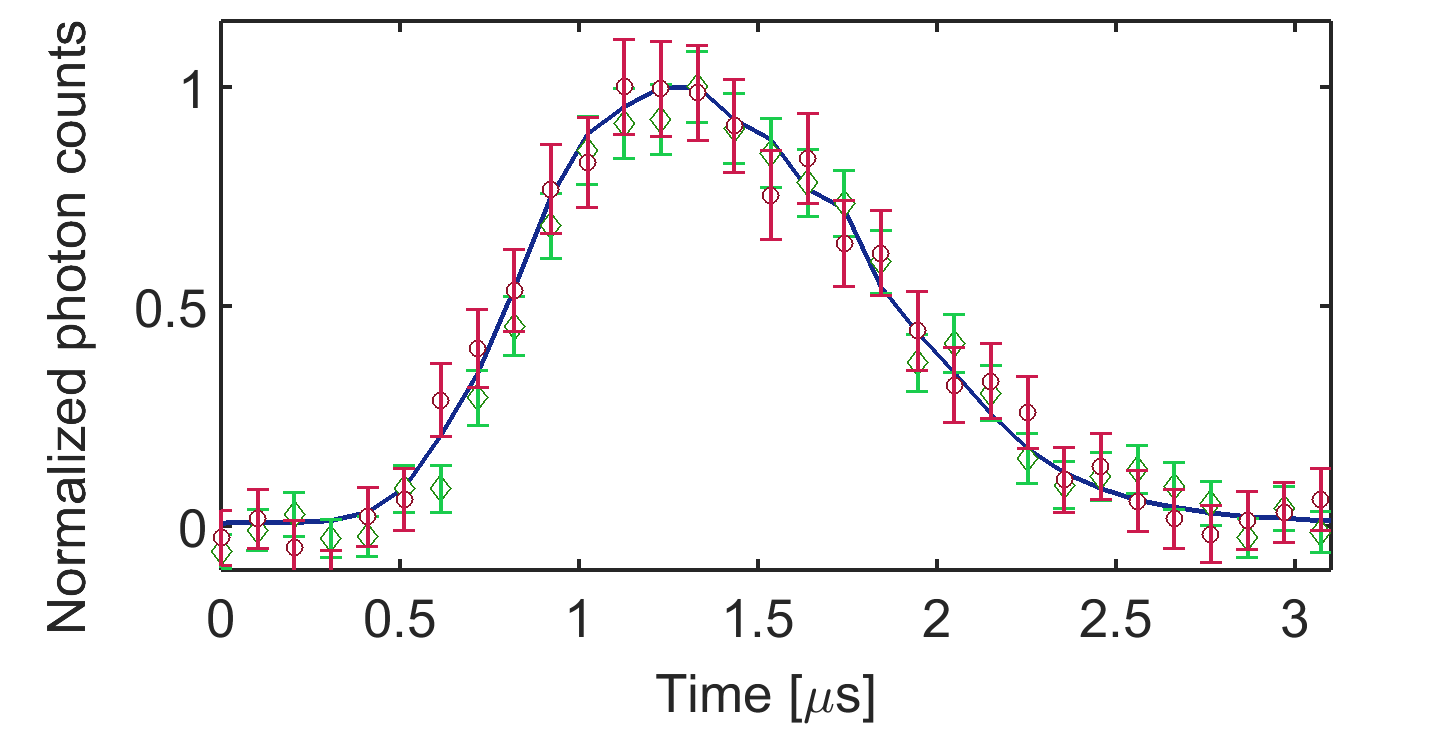}
 \caption{The single photon shape for \SI{866}{\nm}, shown as a solid line, and telecom photons with and without the \SI{10}{\km} fibre, shown as red circles and green squares respectively. Photon arrival times with respect to the sequence trigger are sorted into \SI{80}{\ns} time bins, and the resulting photon shapes are normalised to unity for comparison after subtracting noise. The photons were collected over 1000 seconds for telecom and 90 seconds for \SI{866}{\nm}.
   \label{fig:shape}}
 \end{center}
\end{figure}

In addition to the preservation of the quantum nature of the photon emission of our ion-cavity system in the QFC, the temporal shape of the photons is an important property of deterministic quantum networks \cite{vasilev2010}, enabling the efficient re-absorption of the photon in the receiving ion-cavity system.
To confirm the preservation of the temporal shape of the photon we have recorded the arrival time distribution of the photon at the detector with respect to the trigger signal from the pulse sequence. The histogram of the arrival times represents the temporal shape of the photons emitted by the ion-cavity systems, the frequency converted photons and the frequency converted photons after passing through the \SI{10}{\km} optical fibre in Figure \ref{fig:shape} shows the measured single photon pulse shapes.
Figure \ref{fig:shape} clearly shows the preservation of the temporal shape for the single photons in the conversion process.

\textbf{Conclusion} In this paper we have demonstrated the efficient frequency conversion of single photons from an ion-cavity system and the preservation of the 
quantum nature and the temporal pulse shape in the conversion process. The agreement between the predicted and measured values for $g^{(2)}(0)$ shows that the ion-cavity system is a true single photon source, and that no multi-photon events are introduced by the QFC system, with observed coincidences being due to the background noise. Furthermore, we have shown that these features can be preserved over long distances by employing a \SI{10}{\km} SMF. A possible improvement to this scheme would be to produce photons of a single defined polarisation by driving a Raman transition between specific Zeeman sub-levels \cite{barros2009}. This would approximately double the efficiency of the system, as the polarisation of the photons could be chosen to match the optimum polarisation of the QFC system. Additionally, there is only 20\% transmission through BPF2 after QFC. This could be increased to 90\% using commercially available filters. With these changes, the increased SBR would allow the distance over which the photons are transmitted to be increased to \SI{100}{\km} while still measuring $g^{(2)}(0) = 0.70$, assuming similar attenuation of signal and noise photons. Over such a distance, the $g^{(2)}$ measurement becomes limited mainly by the intrinsic dark count rate of the detectors, roughly \SI{1}{count\per\s}, rather than the background picked up during QFC. The observation of sub-Poissonian statistics at \SI{866}{\nm} over this distance would be infeasible due to the attenuation of the signal, even with a 100\% efficient source. 

During completion of this work we became aware of related experimental work using a trapped ion and a polarisation-independant QFC system, in which entanglement was generated between the ion and frequency converted photons \cite{bock2017}. However, their system does not employ an optical cavity and the converted single photons are at \SI{1310}{\nm}, in the telecom O-band, which is less efficient in optical transmission than the C-band, and is therefore a less viable option for long-distance communication.

The techniques and results in this paper present a wide range of opportunities. The entanglement of trapped ions separated by \SI{1}{\m} has been demonstrated \cite{moehring2007entanglement}; with QFC to telecom wavelengths, this could be extended to many kilometres. Engineering the temporal shape of the photons emitted by the ion-cavity system to match those from another quantum system, and employing QFC to match the wavelengths of the systems, might provide a way to establish entanglement between disparate quantum systems.
With ion-cavity systems providing superior control over the photon emission process as compared to conventional trapped ion techniques, thus making it the prime candidate for networked quantum information systems, the results presented here are a major step towards this application.

\section*{Acknowledgements}
This work was supported by CREST, JST JPMJCR1671; MEXT/JSPS KAKENHI Grant Number JP15H03704, JP16H02214, JP16H01054, JP15KK0164 and JP16K17772; JSPS Bilateral Open Partnership Joint Research Projects. We also gratefully acknowledge support from EPSRC through the UK Quantum Technology Hub: NQIT - Networked Quantum Information Technologies EP/M013243/1 and EP/J003670/1.

\section*{Author Contributions}
T.W., S.V.K. performed the experiment.
T.W. and H.T. analysed the data.
S.V.K. made the figures.
R.I., T.Y., Y.T., K.H. and N.I. planned and designed the QFC system.
K.M., Y.T., R.I. built the QFC system and characterized the performance under
supervision of K.H., T.Y. and N.I.
T.W., M.K., and T.Y. wrote the manuscript. 
All authors contributed to the discussions and interpretations.

\section*{Competing Financial Interests}
The authors declare no competing financial interests.

\end{document}